\input{aipcheck}

\documentclass[
    ,final            
  ]
  {aipproc}

\layoutstyle{6x9}

\begin{document}

\title{Spin fluctuations in cuprates as the key to high $T_c$}


\author{P. Prelov\v sek}{
  address={Faculty of Mathematics and Physics, University of
Ljubljana, Ljubljana, Slovenia}, ,altaddress={J. Stefan Institute,
Ljubljana, Slovenia} }
\author{I. Sega}{
  address={J. Stefan Institute, Ljubljana, Slovenia}
}
\author{A. Ram\v sak}{
  address={Faculty of Mathematics and Physics, University
of Ljubljana, Ljubljana, Slovenia},
,altaddress={J. Stefan Institute, Ljubljana, Slovenia}
 }
\author{J. Bon\v ca}{
  address={Faculty of Mathematics and Physics, University
of Ljubljana, Ljubljana, Slovenia},
,altaddress={J. Stefan Institute, Ljubljana, Slovenia}
}

\begin{abstract}
Spin fluctuations represent the lowest established energy scale in
cuprates and are crucial for the understanding of anomalous normal
state properties and superconductivity in these materials
\cite{imad}. The memory-function approach to the spin response in the
$t$-$J$ model is described. Combined with numerical results for small
systems it is able to explain the anomalous scaling at low doping and
the crossover to the Fermi-liquid-like behavior in overdoped
systems. Within the superconducting phase the theory reproduces the
resonant peak and its peculiar double dispersion. Such spin
fluctuations are then used as the input for the theory of
superconductivity within the $t$-$J$ model, where we show that an
important role is played also by the next-nearest-neighbour hopping
parameter $t'$.
\end{abstract}

\maketitle

\section{Introduction}

The phase diagram of cuprates still represents one of the major
challenges in solid state physics, both for theoreticians and
experimentalists.  Besides superconductivity (SC) and
antiferromagnetic (AFM) ordering, several regimes with distinct
electronic properties have been identified within the normal metallic
phase. In this contribution, we are concerned with the spin dynamical
response in cuprates which has been intensively studied using the
inelastic neutron scattering (INS)
\cite{kast,fong} and NMR relaxation experiments \cite{bert}. 

There is an abundant evidence that in underdoped cuprates magnetic
properties are not following the usual Fermi-liquid (FL) scenario
within the metallic state above the SC transition $T>T_c$. It should
be reminded that within a normal FL the dynamical spin susceptibility
$\chi_{\bf q}^{\prime\prime}(\omega)$ is essentially $T$- independent
at low $T<T_{FL}$. From the point of spin response one can use the
latter criterion as the working definition of the Fermi-liquid
temperature $T_{FL}$. Note that in usual metals $T_{FL} \gg 1000$~K,
in striking contrast to underdoped cuprates with $T_{FL}<T_c$
\cite{kast,bert}. Evidence for the non-Fermi-liquid (NFL) behavior are
INS results for the ${\bf q}$-integrated spin susceptibility which
exhibit in a broad $\omega, T$ range an anomalous, but universal
$\chi_L''(\omega) \propto f(\omega/T)$, established in underdoped
La$_{2-x}$Sr$_x$CuO$_4$ (LSCO) \cite{keim,kast}, as well as in
YBaCu$_3$O$_{6+x}$ (YBCO) \cite{kaku}. Similar conclusions arise from
evident $T$-dependence of $^{63}$Cu NMR spin-lattice relaxation rate
$1/T_1 T$ and of the spin-spin relaxation rate $1/T_{2G}$ \cite{bert}.

Cuprates at optimum doping, and even more in the overdoped regime,
approach closer the usual FL description. INS reveals only weak spin
response at low energies $\omega$ in the normal state (NS) at $T>T_c$,
characteristic for metals with a broader band. Also, NMR relaxation
rates $1/T_1 T$ and $1/T_{2G}$ are weakly $T$-dependent, again
consistent with the FL scenario. Analogous message arises from the
analysis of cuprates doped with nonmagnetic Li and Zn \cite{bobr},
where the impurity-induced spin susceptibility varies as $\propto
1/(T+ T_K)$, whereby the characteristic (Kondo-type) temperature is
$T_K \sim 0$ in the underdoped regime and increasing fast with doping
in the overdoped regime.

While in the NS the dynamical spin response $\chi_{\bf
q}^{\prime\prime}(\omega)$ is in general compatible with an overdamped
collective mode, the prominent feature appearing in the SC phase is
the magnetic resonant mode. First observed in optimally doped YBCO
\cite{ross}, it has been in the last decade the subject of numerous
INS experiments \cite{fong}. The resonant response, in spite of
evident differences between YBCO and LSCO systems, as well as changes
with doping, reveals some surprisingly universal characteristics. The
peak intensity is highest at the commensurate wavevector ${\bf
Q}=(\pi,\pi)$, while its frequency $\omega_r$ increases with doping up
to the optimum doping. In addition, one component of the resonant mode
disperses downwards \cite{bour}, while another branch apparently
emerging from the same peak shows an upward dispersion
\cite{arai,pail,rezn}.   

It seems a plausible (although surprisingly not generally accepted)
conclusion, that the understanding of spin dynamics is the key for the
proper description of the anomalous properties of cuprates, and to the
mechanism of high $T_c$ in particular. The main argument remains, that
up to now the spin fluctuations represent the lowest (experimentally
well established) energy scale in cuprates, both in the NS as well as
in the SC state. Namely, the peak in the spin response (as measured by
INS) in the normal state appears in underdoped cuprates at $\omega_p
\sim T_c$, moving even lower with decreasing $T$. Moreover, the
collective mode $\omega_r$ in the SC phase lies even below the SC gap
$\omega_r<2 \Delta_0$. On the other hand, low spin-fluctuation energy
scale also sets a clear limit to the FL behavior since FL can become
normal only for $T,\omega$ which are below this scale.

A comprehensive theoretical description of spin fluctuations in
cuprates and their implications on other properties, in particular
their role in the mechanism of SC, is still lacking. A FL behavior in
the overdoped regime far from a metal-insulator transition seems
plausible, nevertheless a solid theoretical approach is missing even
in this regime. A crossover from a strange metal to a coherent metal
phase has been predicted within some theoretical approach.  Quite
fashionable and frequently invoked interpretation is given in terms of
the quantum critical point (QCP) at optimum doping $c_h^*$ (masked,
however, at low $T$ by the SC phase), dividing the FL phase at
$c_h>c_h^*$ from a (singular) non-Fermi-liquid (NFL) metal at
$c_h<c_h^*$. Such a scenario is established, e.g. theoretically in
spin systems \cite{sach} and experimentally in some heavy-fermion
compounds, but remains controversial in cuprates. The obvious argument
against the QCP scenario is the absence of a critical length scale,
e.g., AFM correlation length $\xi(T\to 0)\to \infty$ as well as the
absense of the phase with the AFM long-range order for $c_h<c_h^*$
(sometimes put in connection with the pseudogap scale $T^*$). In the
underdoped cuprates the spin fluctuation seem to follow quite well the
phenomenological scenario of the marginal FL \cite{varm}, which got so
far only partially a solid theoretical foundation. The main
distinction to the QCP scenario is the absense of a critical length
scale, which is in agreement with low-energy INS revealing at low $T$
the saturation of the inverse AFM correlation length $\kappa=1/\xi$,
at least in YBCO \cite{fong} and in LSCO at low doping
\cite{keim,kast}.

With respect to the most challenging problem, the mechanism of SC in
cupratres, the role of strong correlations and the antiferromagnetic
(AFM) state of the reference insulating undoped compound has been
recognized very early \cite{ande}. Still, up to date there is no
general consensus whether ingredients as embodied within the prototype
single-band models of strongly correlated electrons are sufficient to
explain the onset of high $T_c$. Even within the frequently invoked
$t$-$J$ model, being the subject of this paper, proposed mechanism of
SC and the methods for the evaluation of correponding $T_c$ differ
with respect to the fact whether the attractive interaction is mainly
local and instantaneous \cite{bask} or the retardation effects are
important \cite{mont}. Recognizing the very low spin-fluctuation scale, we
will advocate in the following the latter for spin-fluctuation
scenario, emerging in contrast to previous approaches directly from
the strongly correlated $t$-$J$ model.

In the following we present some of our recent theoretical results on
spin fluctuations in cuprates and their relation to SC. The analysis
within the $t$-$J$ model, as relevant for cuprates, is mostly based on
the general memory-function approach and the equations-of-motion (EQM)
method, The latter has been first applied to the $t$-$J$ model to
explain anomalous (MFL-type) properties of NS spectral function
\cite{prel} and then extended to low-doping regime \cite{pr1} and SC
\cite{plak}.  Spin dynamical response $\chi_{\bf q}(\omega)$ has been
considered within analogous treatment to yield the overdamped mode in
the NS and resonant peak dispersion in the SC state \cite{spb}, the
anomalous $\omega/T$ scaling in the underdoped regime \cite{psb}, the
influence of nonmagnetic impurities
\cite{ps}, the NFL-FL crossover in spin dynamics \cite{bps}, and 
double dispersion of resonant peak
\cite{sp}. The extracted knowledge on spin fluctuations is used as an
input the theory of SC \cite{pr2}.

\section{NFL - FL crossover in the normal state}

To be specific, we consider in the following the spin dynamics within
the framework of the extended $t$-$J$ model, which has been shown to
represent surprisingly well several electronic properties of cuprates,
both qualitatively and quantitatively \cite{jakl},
\begin{equation}
H=-\sum_{i,j,s}t_{ij} \tilde{c}^\dagger_{js}\tilde{c}_{is}
+J\sum_{\langle ij\rangle}({\bf S}_i\cdot {\bf S}_j-\frac{1}{4}
n_in_j), \label{tj} 
\end{equation}
including in general both the NN hopping $t_{ij}=t$ and the NNN
hopping $t_{ij}=t^\prime$, and involving the projected fermionic
operators, $\tilde{c}_{is}= (1-n_{i,-s}) c_{is}$.

We will first argue \cite{psb} that the anomalous $\omega/T$ scaling
and the related NFL behavior of the magnetic response can be
understood as a consequence of few simple ingredients which appear to
be valid for doped AFM in the normal state: a) the collective mode is
strongly overdamped, whereby the damping is nearly $\omega$- and $T$-
independent at low $\omega$, and b) there is no long-range spin order
at low $T$, so that static spin correlations saturate with a finite
$\xi$.

Within the memory function approach the dynamical spin susceptibility
$\chi_{\bf q}(\omega)=-\langle \!\langle S^z_{\bf q};S^z_{\bf q}
\rangle \!\rangle_{\omega}$ can be generally expressed \cite{spb,psb} 
in the form
\begin{equation}
\chi_{\bf q}(\omega)=\frac{-\eta_{\bf q}}{\omega^2+\omega 
M_{\bf q}(\omega) - \omega^2_{\bf q}}\,, \label{chiq}
\end{equation}
suitable for the analysis of the magnetic response, as present in
undoped and moderately doped AFM \cite{spb}. $\omega_{\bf q}$ represents
the frequency of the  collective mode provided that the mode damping is
small, i.e., $\Lambda_{\bf q}\sim M^{\prime\prime}_{\bf q} (\omega_{\bf
q}) <\omega_{\bf q}$. In the opposite case, i.e. $\Lambda_{\bf
q}>\omega_{\bf q}$ the mode is overdamped. Still, the advantage of the
form (\ref{chiq}) is that it can fullfil basic sum rules even for an
approximate $M^{\prime\prime}_{\bf q}$. Thermodynamic quantitites
entering Eq.~(\ref{chiq}) can be expressed as
\begin{equation}
\eta_{\bf q}=-{\dot\iota}\langle [S^z_{-\bf q}\, 
,\dot{S}^z_{\bf q})]\rangle\,,
\qquad \omega^2_{\bf q}=\eta_{\bf q}/\chi^0_{\bf q} \,,
\end{equation}
where $\chi^0_{\bf q}=\chi_{\bf q}(\omega=0)$ is the static
susceptibility.

$\eta_{\bf q}$ is the spin stiffness and can be expressend in terms of
the static correlation functions, in particular within $t$-$J$ model
$\eta_{\bf Q}= - \langle H \rangle/N$.  $\chi^0_{\bf q}$ (or
$\omega_{\bf q}$) remains to be determined, even for known $M_{\bf
q}(\omega)$.  It is quite a sensitive quantity, hence it safer to fix
it by the sum rule (fluctuation - dissipation relation)
\begin{equation}
\frac{1}{\pi}\int_0^\infty d\omega ~{\rm cth}\frac{\omega}{2T}
\chi^{\prime\prime}_{\bf
q}(\omega)= \langle S^z_{-{\bf q}} S^z_{\bf q}\rangle = C_{\bf q}\, ,
\label{eqsum}
\end{equation}
given in terms of equal time spin correlations, which are expected to
be much more robust. Moreover $C_{\bf q}$ are bound by the constraint
$(1/N)\sum_{\bf q} C_{\bf q} = (1-c_h)/4$, where $c_h$ is an effective
hole doping.

Let us now state two basic assumptions: a) static correlations are
taken to follow a Lorentzian form, i.e. $C_{\bf q}=C/(\kappa^2+\tilde
q^2)$ where $\tilde {\bf q}={\bf q}-{\bf Q}$.  $\kappa$ is assumed to
be a noncritical quantity, which on approaching low $T$ saturates at a
finite value. As already noted, this is consistent with the neutron
scattering data for weakly doped LSCO \cite{keim} and YBCO
\cite{kaku}. It is also consistent with numerical results 
for the $t$-$J$ model at finite doping. b) The damping is also assumed
to be a constant, $M^{\prime\prime}_{\bf q}(\omega) \sim
\Lambda$, i.e., (roughly) independent of $\omega$ , $\tilde {\bf q}$
and $T$, or at least not critically dependent on these variables. The
support for this assumption comes from our numerical results on small
systems, using the finite-$T$ Lanczos method (FTLM) \cite{jakl}, for
the $t$-$J$ model on small lattices with up to 20 sites. Calculating
$\chi^{\prime\prime}_{\bf q}(\omega)$ and extracting then
$M^{\prime\prime}_{\bf q}(\omega)$ with the help of Eq.(\ref{chiq}),
one can conclude \cite{psb} that in spite of widely different
$\chi^{\prime\prime}_{\bf q}(\omega)$ the damping function
$M^{\prime\prime}_{\bf q}(\omega)$ is nearly constant in a broad range
of $\omega<t$ and almost independent of ${\bf q}$. Moreover, for doped
systems with $c_h>0$ data are consistent with a finite (and quite
large) extrapolated value $\Lambda_{\bf Q}(T \to 0)$. In the normal
state this leads to a overdamped collective mode vicinity of ${\bf
q}={\bf Q}$, i.e., $\omega_{\bf Q} < \Lambda$, as generally observed
in INS experiments\cite{imad,kast,fong},
\begin{equation}
\chi''_{\bf q}(\omega) \sim \frac{\eta}{\Lambda} \frac{\omega}
{(\omega^2 + \Gamma^2_{\bf q}) } , \qquad \Gamma_{\bf q}=
\frac{\omega^2_{\bf q}}{\Lambda}, \label{chiim1}
\end{equation}
and $\Gamma_{\bf q}<\omega_{\bf q}$. At low $T \to 0$, Eq.(\ref{eqsum})
leads now to a nontrivial restriction for $\omega_{q}$ and
$\Gamma_{\bf q}$. The relevant quantity is the peak frequency
$\omega_p=\Gamma_{\bf Q}(T\to 0)$, which determines the characteristic
$T=0$ spin-fluctuation scale as well as $T_{FL}$.  

The crucial parameter appears to be
\begin{equation}
\zeta = C\pi\Lambda/(2 \eta \kappa^2), \qquad \omega_p \sim \Lambda 
{\rm e}^{-2\zeta}, \label{zeta}
\end{equation}
which exponentially renormalizes $\omega_p$. Since $C \sim O(1)$ and
$\eta \sim 0.6~t$ at low doping, $\zeta$ is effectively governed by
the ratio $\Lambda/\kappa^2$. It is easy to imagine the situation that
$\zeta \gg 1$ in the underdoped cuprates, leading to very low
$\omega_p \ll \Lambda$ and even $\omega_p < T_c$. On the other hand,
in the overdoped case $\Lambda/\kappa^2 \sim 1$ and $\omega_p$ becomes
large as in usual FL systems. Due to exponential dependence in
Eq.(\ref{zeta}) it is also plausible that the crossover from the NFL
regime with extremely small $\omega_p$ and FL behavior is quite abrupt
\cite{bps}, resembling the QCP scenario.

In order to extract the characteristic energy scale $\omega_{FL}$ of
spin fluctuations directly from numerical FTLM results \cite{bps}, we
use an alternative definition,
\begin{equation}
\omega_{FL}(T)=S_{\bf Q}/\chi_{\bf Q}(T)
\end{equation}
with the corresponding $T=0$ limit $\omega_{FL}(0)$.  Note that
$\omega_{FL}(0)=\langle \omega \rangle$ is just the first frequency
moment of the shape function $\chi_{\bf Q}''(\omega)/\omega$,
\begin{equation}
\omega_{FL}(0)=\langle \omega \rangle=\frac{2}{\pi \chi_{\bf Q}}
\int_0^\infty  \chi_{\bf Q}''(\omega) d\omega.
\end{equation}
On the other hand, one can extract $\omega_{FL}$ also from
experiments, in particular from NMR $1/T_{2G}$ relaxation data
\cite{bert}, which give rather straightforward information on
$\chi_{\bf Q}(T)$.

In Fig.~1 we show FTLM results for $\omega_{FL}(c_h)$ at $0.1 t 
\leq T \leq J$. Besides we also present values extrapolated to $T
\to 0$. Note that in the considered $T$ window $S_{\bf Q}(T)$ is essentially
$T$-independent, following well the linear variation $1/S_{\bf Q}=K
c_h$.  In contrast, the FL scale $\omega_{FL}$ reveals a nonuniform
variation with doping.  Again, for $c_h>c_h^*$, $\omega_{FL}$ is
already rather $T$-independent for $T<J$. On the other hand, in the
regime $c_h<c_h^*$ we find a strong $T$-dependence of $\omega_{FL}$
even at lowest reliable $T$, where $\omega_{FL}
\sim T+ \omega_{FL}(0)$. We can summarize results in Fig.~1 as
follows: a) in the overdoped regime $\omega_{FL}(0) \sim
\alpha (c_h-c_{h0})$ with $c_{h0} \sim 0.12 $ and a large slope
$\alpha \sim 3.5t \sim 1.4$ eV, b) in the underdoped regime our
results indicate on a smooth crossover to very small $\omega_{FL}(0)
\ll J$.

\begin{figure}[htb]
\includegraphics[height=.4\textheight,angle=-90]{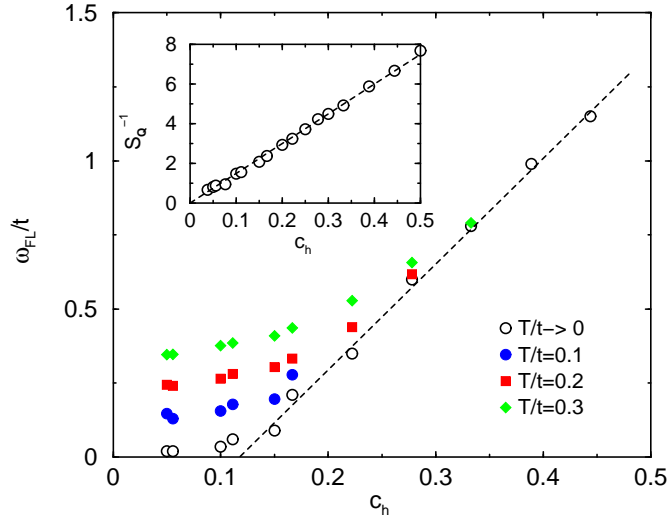} 
\caption{FL scale $\omega_{FL}/t$ vs. $c_h$, obtained
for the $t$-$J$ model using the FTLM for $T>0$ and their $T\to 0$
extrapolated values. The inset shows $T=0$ results for $1/S_{\bf Q}$
vs. $c_h$. Dashed lines are guide to the eye only.}
\label{fig1}
\end{figure} 

Let us estimate $\chi_{\bf Q}(T)$ and consequently $\omega_{FL}$
directly from experiments on cuprates. Within the normal state the
results for the NMR spin-spin relaxation time $T_{2G}$, obtained from
the $^{63}$Cu spin-echo decay, can be related to static $\chi_{\bf q}$
\cite{mmp}.  Assuming that $\chi_{\bf q}$ is peaked at commensurate
${\bf q}={\bf Q}$ and can be described by a Lorentzian form with a
width $\kappa \ll \pi$, one gets a simplified relation
\begin{equation}
\frac{1}{T_{2G}} \sim 0.083 \kappa F({\bf Q}) \chi_{\bf Q}.
\label{eqt2s}
\end{equation}
$1/T_{2G}$ relaxation rates have been measured and summarized in
Ref.~\cite{bert}, i.e., from underdoped to optimally doped YBCO with
$0.63<x<1$, underdoped YBa$_2$Cu$_4$O$_8$, nearly optimum doped
Tl$_2$Ba$_2$Ca$_2$Cu$_3$O$_{10}$ (Tl-2223) and the overdoped
Tl$_2$Ba$_2$CuO$_{6+\delta}$ (Tl-2201), whereby the normalization with
corresponding $F({\bf Q})$ has been already taken into
account. Relevant $\kappa$ is the one appropriate for low-$\omega$
spin dynamics and measured directly by INS. For YBCO data are taken
from Ref.\cite{bala}, which allows us to evaluate $\chi_{\bf Q}(T)$
from Eq.(\ref{eqt2s}). $S_{\bf Q}$ is so far not experimentally
accessible, so we assume here the $t$-$J$ model results to finally
extract corresponding $\omega_{FL}(T)$ as presented in Fig.~2 for
various cuprates.  Derived $\omega_{FL}(0)$ are well in agreement with
model result in Fig.~1, in particular regarding the large slope in the
overdoped regime and a clear change of scale between the underdoped
and overdoped cuprates.

\begin{figure}[htb]
\includegraphics[height=.4\textheight,angle=-90]{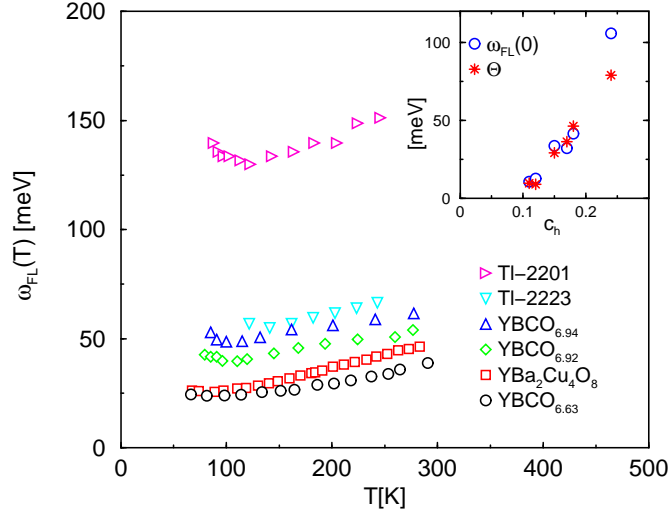} 
\caption{$\omega_{FL}$ vs. $T$, for various cuprates. 
The inset shows the extrapolated scales $\omega_{FL}(0)$ and $\Theta$
(defined by $1/\chi_{\bf Q} \propto T+\Theta$) vs. doping $c_h$.}
\label{fig2}
\end{figure}

\section{Dispersion of the resonant mode}

Using the method of equations of motion within the $t$-$J$ model it
has been shown that the collective spin fluctuations decay into
electron-hole excitations \cite{spb,sp}. This leads to the
lowest-order mode-coupling approximation for the damping in the NS,
\begin{equation}
\Lambda_{\bf q}(\omega) =\frac{\pi}{2\eta_{\bf q}\omega N} 
\int d\omega^\prime [f(\omega^\prime)-f(\omega+\omega^\prime)] 
\sum_{\bf k} w^2_{\bf kq}
A_{\bf k}(\omega^\prime) A_{{\bf k}+{\bf q}}(\omega+\omega^\prime) \, ,
\label{gamma} 
\end{equation} 
where $w_{\bf kq}$ is the effective spin-fermion coupling \cite{spb}
and $A_{\bf k}(\omega)$ is the single-particle spectral
function. Provided the existence of `hot spots' where the FS crosses
the AFM zone boundary (being the case for cuprates at low to
intermediate doping) we assume that in the NS low-$\omega$
quasiparticles (QP) with dispersion $\epsilon_{\bf k}$ and weight $Z_{\bf
k}$ determine the spectral function $A_{\bf k}(\omega)=Z_{\bf k}
\delta(\omega-\epsilon_{\bf k})$. This results in a rather constant
$\Lambda_{\bf q}(\omega)$. The form of Eq.~(\ref{gamma}) is anyhow
quite generic for the damping of the collective magnetic mode in a
metallic system, since the lowest-energy decay processes naturally
involve the electron-hole excitations close to the FS. Similar
expressions appear also in theories based on the RPA approach
\cite{morr,erem}.  Within the SC phase, Eq.~(\ref{gamma}) has
to be generalized to include the anomalous spectral functions
\cite{morr} leading to \cite{spb,sp}
\begin{equation}
\Lambda_{\bf q}(\omega)\sim \frac{\pi}{2 \omega N} \sum_{\bf k}\tilde
w^2_{\bf kq} (u_{\bf k}v_{{\bf k}+{\bf q}}-v_{\bf k}u_{{\bf k}+{\bf
q}})^2 [f(E_{\bf k})-f(E_{\bf k}-\omega)]\,
{\rm\delta}(\omega-E_{\bf k}-E_{{\bf k}+{\bf q}})\bigr] ,
\label{gamsc}
\end{equation}
where $\tilde w^2_{\bf kq}=w^2_{\bf kq}Z_{\bf k}Z_{{\bf k}+{\bf
q}}/\eta_{\bf q}$, while $u_{\bf k},v_{\bf k}$ are the usual BCS
coherence amplitudes and $E_{\bf k}=\sqrt{\epsilon_{\bf
k}^2+\Delta^2_{\bf k}} $.  For the SC gap we assume the $d_{x^2-y^2}$
form, $\Delta_{\bf q}=\Delta_0( \cos q_x -\cos q_y)/2$. Thus we end up
with few adjustable parameters at chosen $c_h$: $\kappa$ in the
Lorentzian form for $C_{\bf q}$, the effective coupling $\bar w$ and
the maximum SC gap $\Delta_0$ \cite{spb,sp}.

\begin{figure}[htb]
\includegraphics[height=.5\textheight,angle=0]{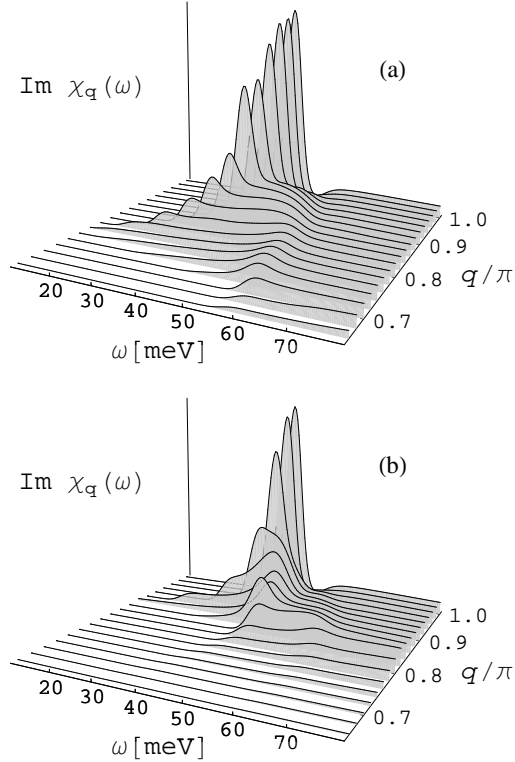} 
\caption{$\chi^{\prime\prime}_{\bf q}(\omega)$
at intermediate doping $c_h=0.15$ for momenta: a) along the $x$
direction ${\bf q}=q(1,0)$, and b) along the zone diagonal ${\bf
q}=q(1,1)$.}
\label{fig3}
\end{figure}
 
At intermediate doping the collective mode is heavily overdamped in
the NS. The indication for the latter is low intensity of the INS in
the relevant low-energy window. For the presented case \cite{sp} we
fix the 'optimum' doping at $c_h=0.15$ and $\kappa\sim 1.25$. The SC
gap is roughly known from experiments $\Delta_0=40$~meV. The remaining
input is $\Gamma_{\bf Q}$ within the NS. For the appearance of the
upper resonant branch it is crucial that $\Gamma_{\bf Q}$ is not too
large, as seems to be inherent within the RPA \cite{morr,erem}.

In Fig.~3 we display $\chi^{\prime\prime}_{\bf q}(\omega)$ for momenta
both along the two ${\bf q}$ directions.  Following observations can
be made \cite{sp}: a) results reveal two branches emerging from the
same coherent resonant mode at $\omega_r \sim 41$~meV. Intensity plots
of both branches within the ${\bf q}$ plane are square-like around AFM
${\bf Q}$, however with quite pronounced anisotropy.  b) For the
downward branch the intensities are strongest along the $(1,0)$
direction. c) The development is more sensitive for $\omega>\omega_r$,
still the situation with the upward branch is just opposite to the
downward one. The dispersion is stronger along the $(1,0)$
direction. d) Above the damping threshold $\omega>~2\Delta_0$ the
upward branch merges into an incoherent response broad both in ${\bf
q}$ as well as in $\omega$. The incoherent part still exhausts most of
the intensity sum rule, Eq.~(\ref{eqsum}), even for ${\bf q}={\bf Q}$.

\section{Spin-fluctuation mechanism of superconductivity}

The projection in fermionic operators in the $t$-$J$ model,
Eq.(\ref{tj}), leads to a nontrivial EQM, which can be in the ${\bf
k}$-basis written as
\begin{equation}
\label{eqm}
[\tilde c_{{\bf k} s},H]= [(1+c_h) \frac{\epsilon^0_{\bf k}}{2} -
(1-c_h)J] \tilde c_{{\bf k} s} + \frac{1}{\sqrt{N}}\sum_{\bf q} m_{\bf
k q} \bigl[ s S^z_{\bf q} \tilde c_{{\bf k}-{\bf q},s} + S^{\mp}_{\bf
q} \tilde c_{{\bf k}-{\bf q},-s} -
\frac{1}{2} \tilde n_{\bf q} \tilde c_{{\bf k}-{\bf q}, s}\bigr],
\end{equation}
where $m_{\bf k q}= 2J \gamma_{\bf q} +
\epsilon^0_{{\bf k}-{\bf q}}$ is the effective spin-fermion coupling, 
while $\epsilon^0_{\bf k}=-4t\gamma_{\bf k}- 4t'\gamma'_{\bf k}$ is
the bare band dispersion on a square lattice, and $\gamma_{\bf
q}=(\mathrm{cos} k_x+\mathrm{cos} k_y)/2, \gamma'_{\bf q}=\mathrm{cos}
k_x \mathrm{cos} k_y$.  To keep similarity with the spin-fermion
phenomenology \cite{mont} we use the symmetrized coupling \cite{spb}
\begin{equation}
\label{mkq}
\tilde m_{\bf kq}= 2J \gamma_{\bf q}+
(\epsilon^0_{{\bf k}-{\bf q}}+\epsilon^0_{\bf k})/2.
\end{equation}

EQM, Eq.~(\ref{eqm}), are used to derive the approximation for the
Green's function (GF) matrix $G_{{\bf k}s}(\omega)= \langle\langle
\Psi_{{\bf k}s}| \Psi^\dagger_{{\bf k}s} \rangle\rangle_\omega$ for
the spinor $\Psi_{{\bf k}s}=(\tilde c_{{\bf k},s},\tilde
c^\dagger_{-{\bf k},-s})$.  We follow the method, as applied to the
normal state (NS) GF by present authors \cite{prel,pr1}, and
generalized to the SC pairing in Ref.\cite{plak,pr2}. In
general, we can represent the GF matrix in the form
\begin{equation}
G_{{\bf k}s}(\omega)^{-1}=\frac{1}{\alpha} [\omega \tau_0 -\hat
\zeta_{{\bf k}s} +\mu \tau_3 -\Sigma_{{\bf k}s}(\omega) ], \label{gf}
\end{equation}
where $\alpha = \sum_i \langle \{\tilde c_{i s},\tilde c^{\dagger}_{i
s}\}_+ \rangle/N = (1+c_h)/2$ is the normalization factor, $\mu$ is
the chemical potential, and the frequency matrix $\hat \zeta_{{\bf
k}s}=\langle \{[\Psi_{{\bf k}s},H], \Psi^{\dagger}_{{\bf k}s} \}_+ 
\rangle/\alpha$, which generates a renormalized band $\tilde \zeta_{\bf k}=
\zeta^{11}_{{\bf k}s}= \bar \zeta - 4 \eta_1 t \gamma_{\bf k}-
4 \eta_2 t' \gamma'_{\bf k}$ and the mean-field SC gap
\begin{equation}
\Delta^0_{\bf k}=\zeta^{12}_{{\bf k}s}= - \frac{4J}{N\alpha} 
\sum_{\bf q} \gamma_{{\bf k} -{\bf q}} \langle 
\tilde c_{-{\bf q},-s} \tilde c_{{\bf q},s} \rangle.
\label{del0}
\end{equation}

To evaluate $\Sigma_{{\bf k}s}(\omega)$ we use the lowest-order
mode-coupling approximation, analogous to the treatment of the SC in
the spin-fermion model \cite{mont}. Taking into account EQM,
Eq.~(\ref{eqm}), and by decoupling fermionic and bosonic degrees of
freedom, one gets
\begin{equation}
\Sigma^{11(12)}_{{\bf k}s}(i \omega_n)=\frac{-3}{N\alpha\beta}
\sum_{{\bf q},m} \tilde m^2_{\bf kq} G^{11(12)}_{{\bf k}-{\bf
q},s}(i \omega_m) \chi_{\bf q}(i \omega_n-i \omega_m) 
\label{sig}
\end{equation}
where $i \omega_n=i \pi(2n+1)/\beta$ and $\beta=1/T$, whereby we have
neglected the charge-fluctuation contribution.

In order to analyze the low-energy behavior in the SC state, we use
the QP approximation for the spectral function matrix where the QP
energies are $E_{\bf k}=(\epsilon^2_{\bf k}+\Delta^2_{{\bf
k}s})^{1/2}$, while NS parameters, i.e., the QP weight $Z_{\bf k}$ and
the QP energy $\epsilon_{\bf k}$, are determined from $G_{{\bf
k}s}(\omega \sim 0)$, Eq.~(\ref{gf}). By defining normalized $F_{\bf
q}(i\omega_l)=\chi_{\bf q}(i\omega_l)/\chi^0_{\bf q}$, and rewriting
the MF gap, Eq.~(\ref{del0}), in terms of the spectral function,
we can display the gap equation in a BCS-like form
form,
\begin{equation}
\Delta_{{\bf k}s}=\frac{1}{N}\sum_{\bf q}[4J \gamma_{{\bf k}-{\bf
q}} -3\tilde m^2_{{\bf k},{\bf k}-{\bf q}}\chi^0_{{\bf k}-{\bf q}}
C_{{\bf q},{\bf k}-{\bf q}}] \frac{Z^0_{\bf k}
Z^0_{\bf q} \Delta_{{\bf q}s}}{2 E_{\bf q}}  \mathrm{th}(
\frac{\beta E_{\bf q}}{2}), \label{del}
\end{equation}
where $C_{\bf k q}=I_{\bf kq}(i\omega_n \sim 0)/I^0_{\bf k}$ plays the
role of the cutoff function with
\begin{equation}
I_{\bf kq}(i\omega_n)=\frac{1}{\beta} \sum_{m} F_{\bf q}(i\omega_n-
i\omega_m) \frac{1}{\omega_m^2 + E^2_{{\bf k}s}}, \label{ikq}
\end{equation}
and $I^0_{\bf k}=\mathrm{th}(\beta E_{\bf k}/2)/(2 E_{\bf k})$.

Eq.~(\ref{del}) represents the BCS-like expression which we use to
evaluate $T_c$. To proceed we need the input of two kinds: a)
$\chi_{\bf q}(\omega)$, and b) the NS QP properties $Z_{\bf
k},\epsilon_{\bf k}$.  As discussed in Sec.~II the NS spin dynamics at
${\bf q} \sim {\bf Q}$ is generally overdamped in the whole doping
regime \cite{fong}. Hence we use the form, Eq.(\ref{chiim1}).  We end
up with parameters $\chi^0_{\bf Q},\Gamma_{\bf Q},\kappa$, which are
dependent on $c_h$, but in general as well vary with $T$. Although one
can attempt to calculate them as described in Sec.~III \cite{spb}, we
use here the experimental input for cuprates, as discussed in Sec.~II
\cite{bps}. For the NS $A_{\bf k}(\omega)$ and corresponding 
$Z_{\bf k},\epsilon_{\bf k}$ we solve Eq.~(\ref{sig}) for
$\Sigma^{11}_{\bf k}=\Sigma_{\bf k}$ as in Ref.~\cite{pr1}, with the
same input for $\chi_{\bf q}(\omega)$. The main message remains
\cite{pr2} that soft AFM fluctuations with ${\bf q} \sim {\bf Q}$ lead
through Eq.~(\ref{sig}) to a reduction of $Z_{\bf k}$, which is ${\bf
k}$-dependent. A pseudogap appears along the AFM zone boundary and the
FS is effectively truncated in the underdoped regime with $Z_{{\bf
k}_F} \ll 1$ near the saddle points $(\pi, 0)$ (in the antinodal part
of the FS) \cite{pr1,pr2}.

Close to half-filling and for $\chi^0_{\bf q}$ peaked at ${\bf q}\sim
{\bf Q}$ both terms in the gap equation, Eq.~(\ref{del}), favor the
$d_{x^2-y^2}$ SC. The mean-field part $\Delta^0_{\bf k}$,
Eq.~(\ref{del0}), involves only $J$ which induces a nonretarded local
attraction, playing the major role in the RVB-type theories
\cite{ande,bask}. In contrast, the spin-fluctuation part represents a
retarded interaction due to the cutoff function $C_{{\bf k}{\bf q}}$,
determined by $\Gamma_{{\bf k}-{\bf q}}$. The largest contribution to
the SC pairing naturally arises from the antinodal part of the Fermi
surface. Meanwhile, in the same region also $Z_{\bf k}$ is smallest
thus reducing the pairing strength, in particular in the underdoped
regime.

One can question the relative role of the hopping parameters
$t,t^\prime$ and the exchange $J$ in the coupling,
Eq.~(\ref{mkq}). While our derivation within the $t$-$J$ model is
straightforward, an analogous analysis within the Hubbard model using
the projections to the lower and the upper Hubbard band, respectively,
would not yield the $J$ term within the lowest order since $J \propto
t^2$. This stimulates us to investigate in the following also
separately the role of $J$ term in Eq.~(\ref{del}), both through the
MF term, Eq.~(\ref{del0}), and the coupling $\tilde m_{\bf kq}$,
Eq.~(\ref{mkq}).

\begin{figure}[htb]
\includegraphics[height=.3\textheight]{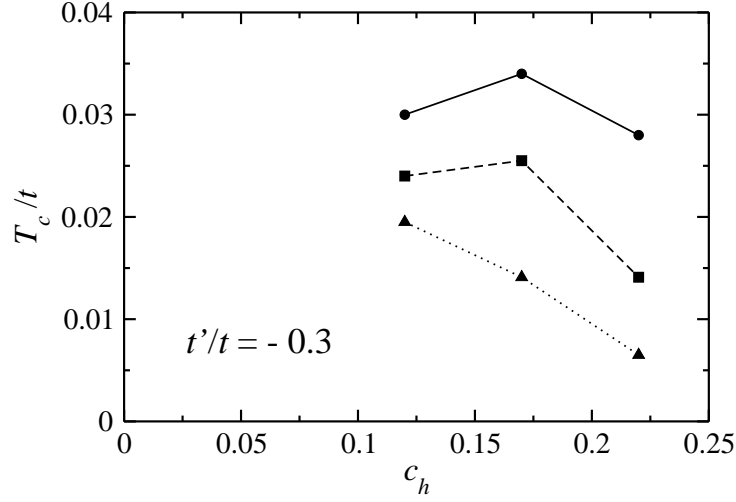} 
\caption{$T_c/t$ vs. doping $c_h$ for $t'/t=-0.3$, calculated for
various versions of Eq.~(\ref{del}): a) full result (full line), b)
with neglected MF term (dashed line), and c) in addition to b)
modified $\tilde m_{\bf kq}$ without the $J$ term (dotted line).}
\label{fig4}
\end{figure}

\begin{figure}[htb]
\includegraphics[height=.3\textheight]{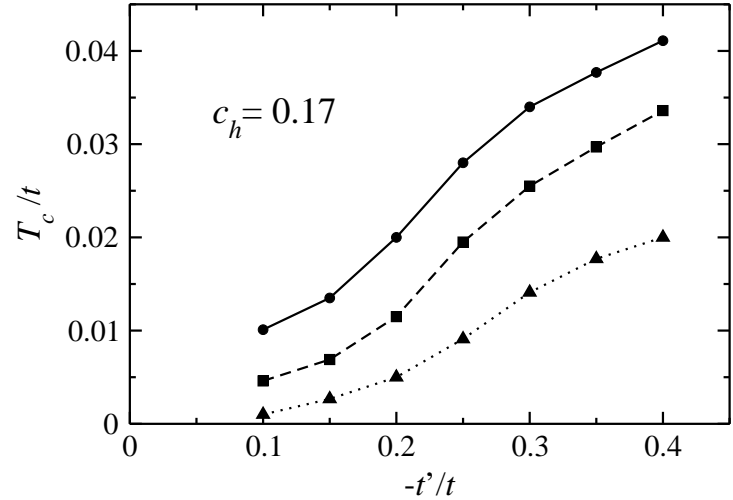} 
\caption{$T_c/t$ vs. $-t'/t$ for fixed 'optimum' doping $c_h=0.17$ and
different versions of Eq.~(\ref{del}), as in Fig.~2.}
\label{fig5}
\end{figure}

Results for the NS spectral properties reveal that the coupling to AFM
fluctuations partly change the shape of the Fermi surface, more
pronounced is however the effect on the QP weight. $Z_{\bf k}$ is
reduced along the AFM zone boundary away from the nodal points.  In
Fig.~4 we present final results for $T_c$, as they follow from the gap
equation, Eq.~(\ref{del}).  It is evident that the spin-fluctuation
contribution is dominant over the mean-field term. When discussing the
role of the $J$ term in the coupling, Eq.~(\ref{mkq}), we note that in
the most relevant region, i.e., along the AFM zone boundary $\tilde
m_{\bf kQ}=2J-4t^\prime \cos^2 k_x$. Thus, for hole doped cuprates,
$t^\prime<0$ term and $J$ term enhance each other in the coupling, and
neglecting $J$ in $\tilde m_{\bf kq}$ reduces $T_c$, although at the
same time relevant $Z_{\bf k}$ is enhanced.

Finally, in Fig.~5 we present results, as obtained for fixed
intermediate doping $c_h=0.17$, but different $t'/t<0$, as relevant
for hole-doped cuprates \cite{pava}. As expected the dependence on
$t'$ is pronounced, since the latter enters directly the coupling
$\tilde m_{\bf kQ}$, Eq.(\ref{mkq}).  It is instructive to find an
approximate BCS-like formula which simulates our results. The latter
involves the characteristic cut-off energy $\Gamma_{\bf Q}$, while
other relevant quantities are the electron density of states ${\cal
N}_0$ and $Z_m$ being the minimum $Z_{\bf k}$ on the FS (in the
antinodal point). Then, we get a reasonable fit to our numerical
results with the expression,
\begin{equation}
T_c \sim 0.5 \Gamma_{\bf Q} ~{\rm e}^{-2/( {\cal N}_0 V_{eff}) },
\label{tc}
\end{equation}
where the effective interaction is given by $V_{eff}= 3 Z_m
(2J-4t^\prime)^2 \chi_{\bf Q}$. 

Probably the most interesting novel result on SC is a pronounced
dependence of $T_c$ on $t^\prime$ which is also consistent with the
evidence from different families of cuprates \cite{pava}. One can give
a plausible explanation of this effect. In contrast to NN hopping $t$,
the NNN $t^\prime$ represents the hopping within the same AFM
sublattice, consequently in a double unit cell fermions couple
directly to low-frequency AFM paramagnons.  Calculated $T_c$ are in a
reasonable range of values in cuprates. We also note that rather
modest 'optimum' $T_c$ value within presented spin-fluctuation
scenario emerge due to two competing effects in
Eqs.~(\ref{del}),(\ref{tc}): large $\tilde m_{\bf kq}$ and $\chi_{\bf
Q}$ enhance pairing, while at the same time through a reduced $Z_{\bf
k}$ and cutoff $\Gamma_{\bf Q}$ they limit $T_c$.  At the same time,
INS experiments \cite{fong} reveal that in underdoped cuprates the
resonant peak at $\omega \sim \omega_r$ takes the dominant part of
intensity of ${\bf q} \sim {\bf Q}$ mode which becomes underdamped
possibly even for $T>T_c$. Thus it is tempting to relate $\Gamma_{\bf
Q}$ to $\omega_r$ and to claim $T_c \sim a \omega_r$, as indeed
observed in cuprates \cite{fong}.



\bibliographystyle{aipproc}   


\end{document}